\documentclass[%
apl,
amsmath,amssymb,
reprint,%
]{revtex4-2}

\usepackage{graphicx}
\usepackage{dcolumn}
\usepackage{bm}
\usepackage[T1]{fontenc}
\usepackage{textcomp}

\usepackage{mathptmx}
\usepackage{units}
\usepackage{amsmath}
\usepackage{sidecap}
\usepackage[labelformat=empty]{subfig}
\usepackage{upgreek}
\usepackage{hyperref}
\usepackage{breakurl}

\begin{document}
	\preprint{AIP/123-QED}
	\setlength{\parindent}{0pt}
	
	\title[Fully Resonant Magneto-elastic Spin-wave Excitation by Surface Acoustic Waves under Conservation of Energy and Linear Momentum]{Fully Resonant Magneto-elastic Spin-wave Excitation by Surface Acoustic Waves under Conservation of Energy and Linear Momentum}
	
	\author{Moritz Geilen}
	\email{mgeilen@physik.uni-kl.de}
	\affiliation{Fachbereich Physik and Landesforschungszentrum OPTIMAS, Technische Universit\"at Kaiserslautern, Germany}
	\author{Alexandra Nicoloiu}
	\affiliation{National Institute for Research and Development in Microtechnologies, Bucharest R-07719, Romania}
	\author{Daniele Narducci}
	\affiliation{imec, Leuven B-3001, Belgium}
	\affiliation{KU Leuven, Departement Materiaalkunde, 3001 Leuven, Belgium}
	\author{Morteza Mohseni}
	\affiliation{Fachbereich Physik and Landesforschungszentrum OPTIMAS, Technische Universit\"at Kaiserslautern, Germany}
	\author{Moritz Bechberger}
	\author{Milan Ender}
	\affiliation{Fachbereich Physik and Landesforschungszentrum OPTIMAS, Technische Universit\"at Kaiserslautern, Germany}
	\author{Florin Ciubotaru}
	\affiliation{imec, Leuven B-3001, Belgium}
	\author{Burkard Hillebrands}
	\affiliation{Fachbereich Physik and Landesforschungszentrum OPTIMAS, Technische Universit\"at Kaiserslautern, Germany}
	\author{Alexandru M\"uller}
	\affiliation{National Institute for Research and Development in Microtechnologies, Bucharest R-07719, Romania}
	\author{Christoph Adelmann}
	\affiliation{imec, Leuven B-3001, Belgium}
	\author{Philipp Pirro}
	\affiliation{Fachbereich Physik and Landesforschungszentrum OPTIMAS, Technische Universit\"at Kaiserslautern, Germany}
	
	\date{\today}
	
	\begin{abstract}
		We report on the resonant excitation of spin waves in micro-structured magnetic thin films by surface acoustic waves (SAWs). The spin waves as well as the acoustic waves are studied by micro-focused Brillouin light scattering spectroscopy. Besides the excitation of the ferromagnetic resonance, a process which does not fulfill momentum conservation, also the excitation of finite-wavelength spin waves can be observed at low magnetic fields. Using  micromagnetic simulations, we verify that during this excitation both energy and linear momentum are conserved and fully transferred from the SAW to the spin wave. 
	\end{abstract}
	
	\maketitle
	
	The rapid increase in computing power in CMOS-based devices, which is described by Moore’s Law, has slowed down in recent years \cite{Moore1965}. One reason for this is the drastic reduction in component size combined with the Joule heating generated by the charge carriers. A promising complementary technology is magnonics \cite{Serga2010,Khitun2010,Chumak2015,Kruglyak2010}, in which spin waves, whose quasi-particles are known as magnons, are used as information carriers. Unlike CMOS, magnonics proposes a wave-based logic in which the information can be encoded in both the amplitude and the phase of the spin waves \cite{Pirro2021}. This also makes magnonic systems attractive for implementing neural networks \cite{Papp2021}. Over time, the functionality of various essential components, such as transistors \cite{Chumak2014}, diodes \cite{Bracher2017,Grassi2020,Szulc2020}, majority gates \cite{Klingler2014,Fischer2017,Talmelli2020}, directional couplers \cite{Sadovnikov2015} or half-adders \cite{Qi2020} was demonstrated. Due to the small wavelength and the GHz frequencies of  spin waves, the necessary miniaturization of the magnonic components to the nanometer scale\cite{Mahmoud2020,Pirro2021} is possible.  However, the efficient excitation of spin waves at these dimensions is challenging, as the usually used micro-antennas through which a microwave current flows suffer from Joule heating, as in the case of CMOS devices, which results in similar problems. This argues for purely voltage-based excitation mechanisms, such as the use of multi-ferroic heterostructures \cite{Fiebig2005,Smolenskii1982,Shelukhin2020}. An alternative approach is to use magneto-elastic coupling between magnons and phonons \cite{Khitun2011a}. In piezoelectric materials, phonons can be excited in an energy-efficient way by means of electric fields. In this context, the coupling of spin waves with bulk acoustic waves (BAW) \cite{Alekseev2020} and with surface acoustic waves (SAWs) is of great interest \cite{Dreher2012,Kuss2021}. It has been demonstrated that the excitation of spin waves by SAWs provides a damping channel for the latter. This has been used to realize non-reciprocal SAW propagation using the Dzyaloshinskii-Moriya interaction \cite{Verba2018,Ku2020}. Most of these studies use SAW transmission spectroscopy \cite{Xu2020}, while only a few take an optical approach to probe the phonon-magnon coupling \cite{Babu2020,Zhao2021,Kraimia2020}. Similarly, most works concentrate on the SAW transmission and not on the properties of the excited spin waves.\\ \noindent
	In this study, we investigate the excitation of spin waves by SAWs in micrometer-sized strips of cobalt-iron-boron (CoFeB) thin films. We use micro-focused Brillouin light scattering spectroscopy (µBLS) to observe both SAWs and spin waves separately.This is possible due to the different rotation of the polarization of the light during the inelastic scattering process. We demonstrate that for a given SAW frequency $f$, resonances for the excitation of spin waves at the same frequency can be observed at two different magnetic bias fields. The excitation at higher fields can be identified as the acoustically driven ferromagnetic resonance (FMR). In this process, however, the wave vector is not preserved in the phonon-magnon scattering process. The second excitation at low bias fields corresponds to a wave vector-preserving excitation of spin waves. To illustrate the details of this process we use micromagnetic simulations which take into account the finite size of the magnetic elements as well as the (inhomogeneous) demagnetization fields inside them.
	
	\begin{figure}[h]
		\includegraphics[width=10cm]{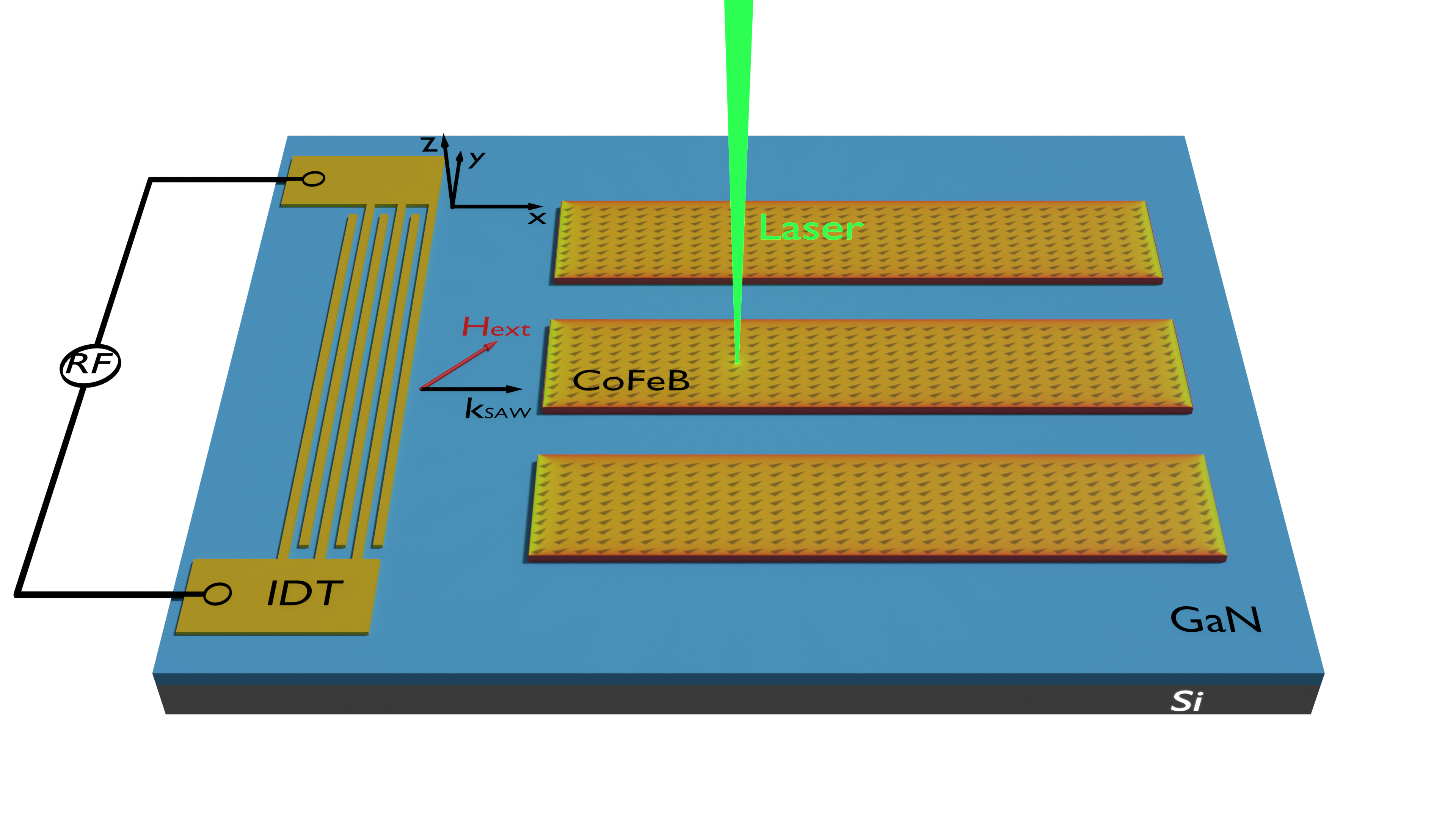}
		\centering
		\caption{Schematic illustration of the investigated structure (not to scale). SAWs are excited by interdigital transducers (IDTs) with $\unit[170]{nm}$ finger-to-finger spacing on a piezoelectric GaN layer. The CoFeB rectangles are $\unit[10]{\upmu m}$ wide, $\unit[50]{\upmu m}$ long and $\unit[18]{nm}$ thick. The external bias field is applied with an angle of $45$° to the SAW propagation direction which is equivalent to the long axis of the CoFeB rectangles.}
		\label{fig:fig1}
	\end{figure}
	\noindent
	Brillouin light scattering \cite{Sebastian2015} is the inelastic scattering of photons with quasi-particles, like magnons or phonons, which makes it a favorable tool to probe the phonon-magnon interaction. During the scattering process energy and momentum of the scattered quasi-particle is transferred to the photon. In this study a single-mode laser with a wavelength of $\uplambda_\mathrm{Laser}=\unit[512]{nm}$ is used. To obtain a sufficient spatial resolution the light is focused onto the sample by a microscope objective with a magnification of 100x and a numerical aperture of NA=0.75. This results in spatial resolution down to $\unit[250]{nm}$ and a maximal detectable wave vector of $k_{\mathrm{max}}=\unit[16]{rad/\upmu m}$. The backscattered light is then collected by the same objective lens and guided to the interferometer. The frequency shift of the scattered light is analyzed by a polarization-sensitive Tandem-Fabry-Perot interferometer (TFPI). 
	The investigated sample is shown in Fig. \ref{fig:fig1} and consists of an interdigital transducer (IDT), which has been fabricated on commercial GaN/Si wafer (produced by NTT-AT Japan). Undoped GaN ($\unit[1]{\upmu m}$) is grown on a Si substrate with a $\unit[0.3]{\upmu m}$ buffer layer. To excite SAWs an IDT with a finger-to-finger distance of $\unit[170]{nm}$ is used, which results in an excited SAW wavelength of $\uplambda_\mathrm{SAW}=\unit[680]{nm}$ ($k_\mathrm{SAW}=\unit[9.2]{rad/ \upmu m}$) at the first harmonic resonance frequency of $f=\unit[6.3]{GHz}$. The IDT has been contacted by microwave probes with ground-signal-ground configuration and connected to a RF-generator. Further details about the phonon excitation and propagation in this particular structure are published in Ref. \cite{Geilen2020}. In this system, both the Rayleigh mode and the first Sezawa mode are excited simultaneously. For these two modes the displacement takes place exclusively in the x-z plane. Consequently, only the components $S_{xx}$ and $S_{xz}$ of the strain tensor have to be considered in the following. As shown in Fig. \ref{fig:fig1}, rectangles of CoFeB are located in front of the IDT. All are $\unit[50]{\upmu m}$ long, $\unit[10]{\upmu m}$ wide and have a thickness of $\unit[18]{nm}$. Additionally, a single microstrip antenna (not shown in Fig. \ref{fig:fig1}) with a width of $\unit[1]{\upmu m}$ is structured over the magnetic rectangles  which serves as a reference spin-wave source. The external magnetic field is applied in-plane and under an angle of $\varphi=$45° to the wave vector of the SAWs. Under this condition, the magneto-elastic interaction is maximized \cite{Dreher2012} and a strong magneto-elastic spin-wave excitation by SAWs is expected.
	
\begin{figure*}[ht]
		\includegraphics[width=16cm]{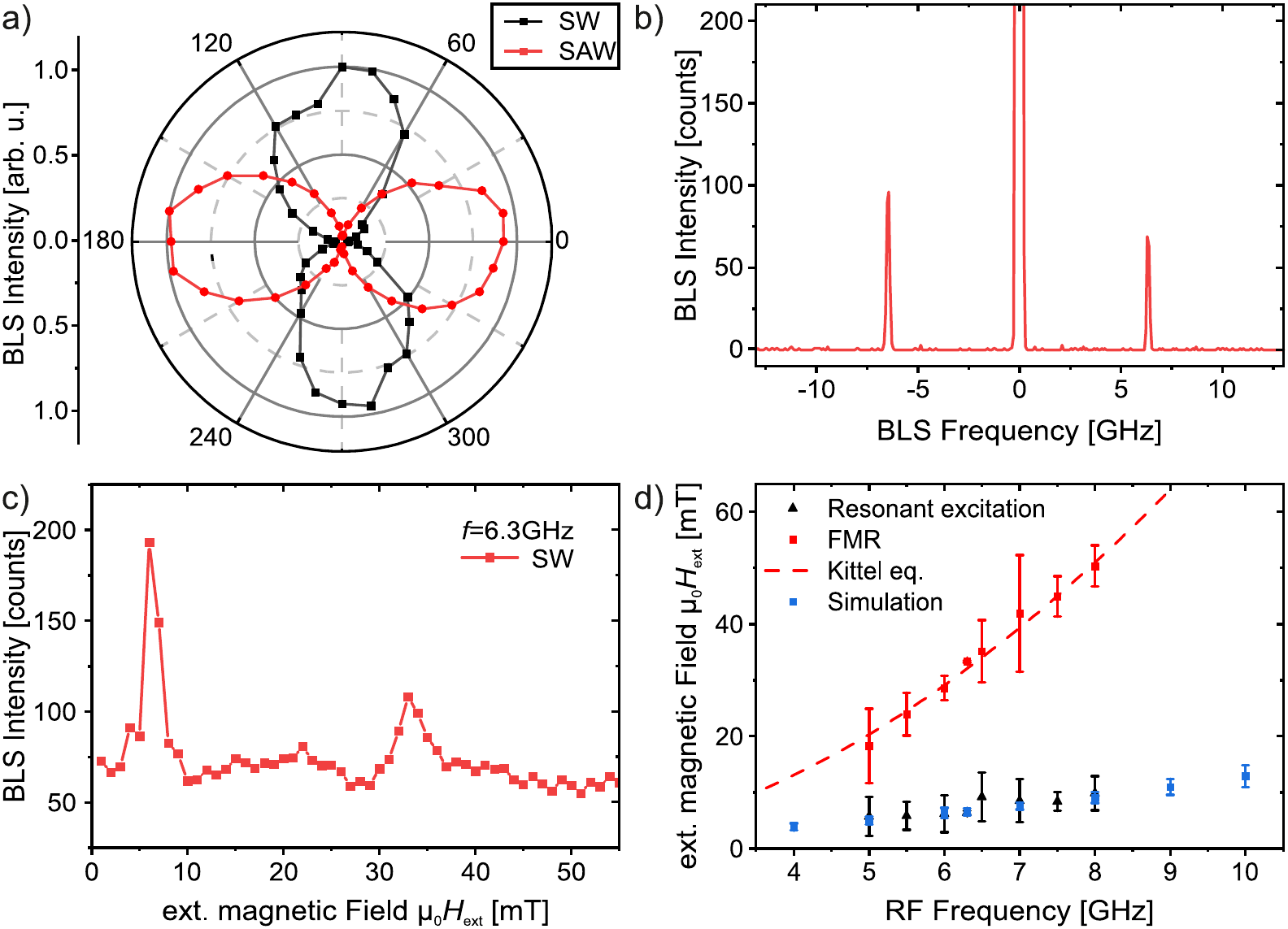}
		\centering
		\caption{a) normalized BLS intensity as a function of the polarization. The polarization of the light scattered by magnons is rotated by $\pi/2$, while the light scattered by SAWs is unchanged in polarization. b) BLS spectrum from spin waves excited by SAWs at a frequency of $f=\unit[6.3]{GHz}$ and an external field of $\upmu_0H_{ext}=\unit[6]{mT}$  c) BLS intensity as a function of the external magnetic field. Two peaks can be identified corresponding to FMR (higher fields) and to wave vector preserving excitation (lower fields). d) Extracted field values from $\upmu$BLS measurements for FMR (red) and resonant excitation (black). FMR data is fitted by Kittel equation. The field values for the resonant excitation from micromagnetic simulations are shown in blue.}
		\label{fig:fig2}
	\end{figure*}

	First, the polarization of the inelastically scattered light is investigated in order to separate the signals originating from spin waves and SAWs. We use a $\uplambda/2$-plate in front of the polarization-sensitive TFPI to analyze the polarization of the inelastically scattered light. For this purpose, SAWs with the frequency $f=\unit[6.3]{GHz}$ were excited by means of the IDT. The measurement takes place on the CoFeB while a strong external magnetic field is applied which shifts the magnon band above the SAW frequency. Figure \ref{fig:fig2} a) shows the normalize BLS intensity as a function of polarization of the inelastically scattered light. For the SAWs (red), the polarization of the scattered light does not change with respect to the incident polarization. To generate a signal which is purely produced by scattering with spin waves, the reference microstrip antenna \cite{Ciubotaru2016} and a microwave current with the same frequency and power is used. As expected, the polarization of the inelastically scattered light is rotated by $\pi/2$. Therefore, the intensity of magnons and phonons can be measured separately by selecting the appropriate polarizer setting.\\
	In the following, the spin waves excited by SAWs are studied. For this purpose, a microwave  with a frequency of $f=\unit[6.3]{GHz}$ and a power of $P=\unit[+5]{dBm}$ (neglecting losses in connecting cables and connections) is applied to the IDT. A typical BLS spectrum is shown in Fig. \ref{fig:fig2} b). It can be seen that only spin waves around the RF-frequency are excited. The measurement of the magnons is carried out close to the edge of the CoFeB rectangle facing the IDT. Only the portion of the light rotated by $\pi/2$ is analyzed. Figure \ref{fig:fig2} c) shows the extracted BLS intensity around $f=\unit[6.3]{GHz}$ for different external magnetic fields. Two maxima can be identified near $\upmu_0 H_{\mathrm{ext}} = \unit[6]{mT}$ and $\upmu_0 H_{\mathrm{ext}} = \unit[33]{mT}$, respectively. The maximum at the higher field values can be attributed to the excitation of the FMR \cite{Duquesne2019}. Here, the energy conservation is fulfilled while the momentum conservation is broken. The second maximum is connected to a phase-matching excitation of spin waves where both energy and linear momentum are conserved as will be discussed further below in detail.
	
	\noindent
	Since the IDTs have a finite length, it is also possible to excite SAWs away from the IDT resonance at $\unit[6.3]{GHz}$. In this case, the excited wave vector of the waves adapts to the dispersion of the excited frequency and the intensity of the SAWs decreases drastically \cite{Geilen2020}. However, since the scattering process is a linear process, qualitatively similar results can be observed for frequencies not coinciding with the IDT resonance. Figure \ref{fig:fig2} d) shows the bias field values of the intensity maxima for different excitation frequencies. The maxima at higher field values (red) correspond to the acoustically driven FMR \cite{Dreher2012} and can be described very well by the Kittel formula \cite{Herring1951}:
	
	\begin{equation}
		f = \gamma \upmu_0 / {2 \pi} \sqrt{(H_\mathrm{eff}+H_\mathrm{k}\cos(\varphi))(H_\mathrm{eff}+H_\mathrm{k}\cos(\varphi)+M_\mathrm{S})}
	\end{equation} 
	\noindent
	A saturation magnetization of $M_S=\unit[1150]{kA/m}$ and a gyromagnetic ratio of $\gamma=\unit[182]{rad/{ns T}}$ are found. Furthermore, a small unidirectional anisotropy field $H_\mathrm{k}=\unit[2.91]{mT}$ pointing along the x-axis is included (see Appendix). 
	For the spin waves excited at low fields, only a slight increase in resonance field with the excitation frequency is visible. In the following, using micromagnetic simulations we will illustrate that this excitation is the result of a direct conversion of a phonon into a magnon, in which both the frequency and the wave vector is preserved. Since the wave vector of the SAW increases almost linearly with its frequency, the resonance field for the resonant scattering to the spin wave of the same frequency is increasing only slightly, in contrast to the non-resonant excitation of the FMR. \\
	To verify our hypothesis, we conduct micromagnetic simulations using Mumax3 \cite{Vansteenkiste2014}. The software platform Aithericon was used to automatically start and analyze the simulation series and to manage the produced data \cite{aithericon}. The parameters used for the simulation are as follows: $M_\mathrm{S}=\unit[1150]{kA/m}$, $A_{\mathrm{ex}}=\unit[15]{pJ/m}$, $B_1=B_2=\unit[-8.8]{MJ/m^3}$ \cite{Vanderveken2021,Gueye2016} and $\alpha=0.0043$. For the uniaxial in-plain anisotropy along the x-axis, an anisotropy constant of $K_\mathrm{u}=\unit[1600]{J/m^3}$ was used (see Appendix). The dimensions of the simulated rectangle corresponds to the experimental ones. The simulated volume is divided into 2560x512x1 cells. To mimic the SAWs we assume a plane wave for the strain components $S_{xx}$ and $S_{xz}$ with wave vector $k_{\mathrm{SAW}}$ and frequency $f$. The ratio of $S_{xx}$ and $S_{xz}$ is unity and their phase is shifted by $\pi/2$, which results in the approximation of circular motion. Mumax3 calculates the effective magnetic field resulting from the SAW (see Appendix). The amplitude of the acoustic wave is chosen small enough such that non-linear effects in the spin-wave system can be neglected. 
	
	\begin{figure*}[ht]
		\includegraphics[width=16cm]{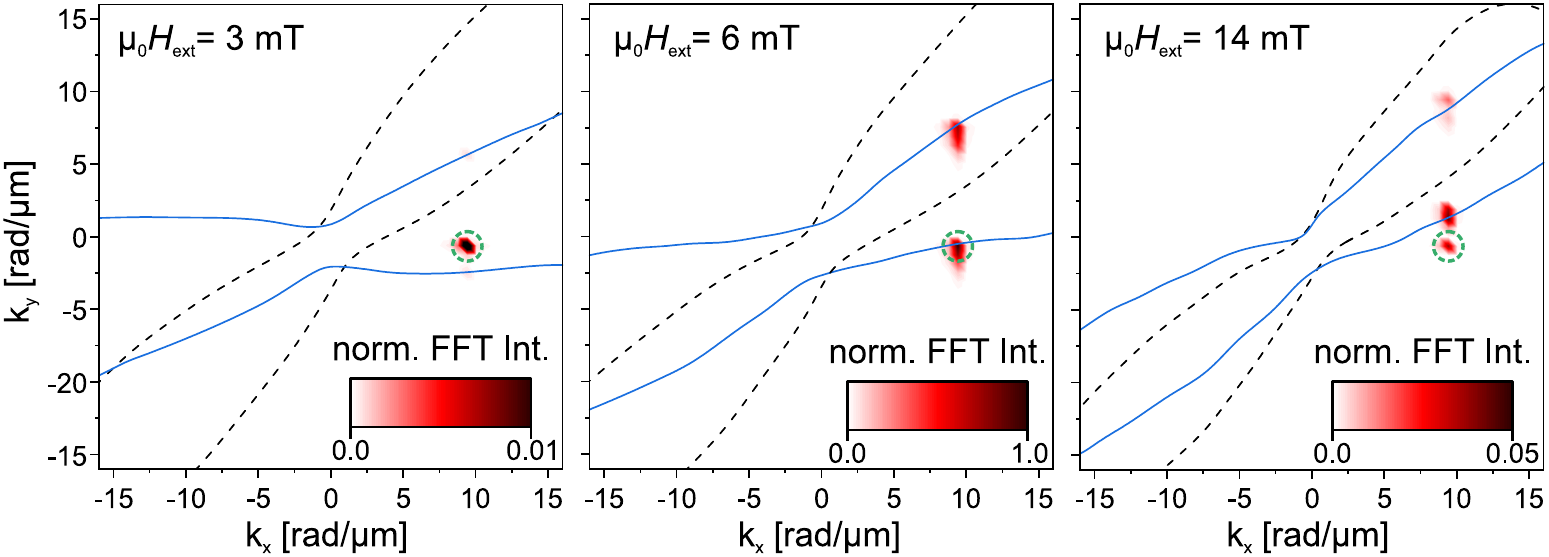}
		\centering
		\caption{Normalized spin-wave (FFT) intensity in reciprocal space for different external magnetic fields extracted for the SAW frequency $f_\mathrm{SAW}=\unit[6.3]{GHz}$. The isofrequency curves for the investigated structure (obtained from micromagnetic simulations using a point-like spin wave source) and an infinitely extended film are shown in solid blue and dashed black lines, respectively. The excitation with the wave vector of the SAW, which is only resonant for $\mu_0 H_\mathrm{ext} =$6 mT, is indicated by green circles.}
		\label{fig:fig3}
	\end{figure*}
\begin{figure}[h]
		\includegraphics[width=8cm]{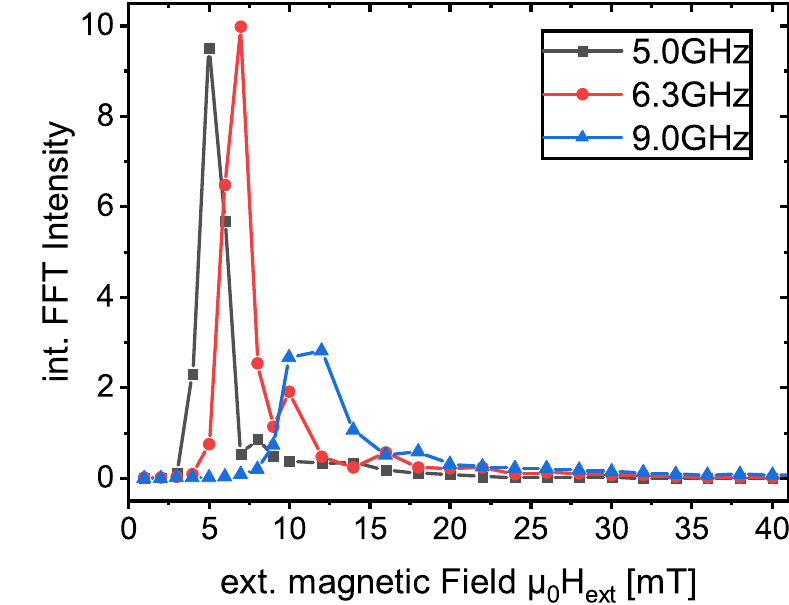}
		\caption{Integrated spin-wave intensity from micromagnetic simulations for different SAW frequencies as a function of the external field. The peaks are associated to the fully resonant excitation of spin waves by the SAWs.  }
		\label{fig:fig4}
	\end{figure}

	\noindent
	By performing a Fourier transform of the dynamic magnetization components in time and space we obtain the spin wave intensity in $k$-space. To determine which wave vectors are excited, Fig. \ref{fig:fig3} shows in color-code the simulated spin-wave intensity in reciprocal space exemplary for the frequency $f_{\mathrm{SAW}} = \unit[6.3]{GHz}$. It can be seen that the magneto-elastic field generated by the SAWs drives the magnetization for all fields with the wave vector of the SAWs, which is indicated by green circles. However, this excitation is not necessarily an eigenstate of the magnonic system, and relaxation into states on the isofrequency curve (illustrated in blue) occurs. During the relaxation, the wave vector along the propagation direction of the SAWs (x-axis) is preserved. Most importantly, at $\upmu _0 H_{\mathrm{ext}}=\unit[6]{mT}$, it can be seen that the direct excitation by the magneto-elastic field coincides with the magnonic isofrequency curve. This results in a resonant excitation which shows a drastically increased amplitude (note the different scales), in good agreement with the experimental findings in Fig. \ref{fig:fig2} c). 
	To obtain a quantitative picture, we compare the excited spin-wave intensity in the simulations with our experimental results. Since $\upmu$-BLS collects all spin waves up to  maximum wave vector $k_\mathrm{max}=\unit[16]{rad/\upmu m}$ also the simulated spin-wave intensities are integrated up to this value. The obtained values are shown in Fig. \ref{fig:fig4} for different external fields. All curves show a clear maximum at low magnetic fields, which corresponds to the resonant excitation. The extracted values are shown in blue in Fig. \ref{fig:fig2} d) and match  well to the measured values of the resonant excitation. However, the excitation of the FMR is not significantly visible in the simulations. A closer analysis of the micromagnetic simulations shows only that the intensity for fields above the FMR decreases drastically, since in this region only the excitation of edge modes is possible. However, there is no intensity peak associated with the FMR in the simulations. There are several possible explanations for this absence of the FMR excitation in the micromagnetic simulations. First, no inhomogeneities are considered in the simulation, which would break the wave vector conservation rule in the scattering process. Second, only the Rayleigh mode is considered in the simulations. To what extent the interference between Rayleigh- and Sezawa mode and the associated interference pattern\cite{Geilen2020} can also contribute to the excitation of the FMR is still part of further investigations. Most importantly, further investigations on extended films have shown that the resonance attributed to the FMR cannot be observed there. This suggests that the edges of the magnetic structure significantly contribute to the FMR excitation in the small structures presented here.
	\noindent
	We would like to note that in the particular case of this study, the resonant excitation is possible due to the anisotropy present in the magnetic structures. This anisotropy is composed of the shape anisotropy of the micro-structure and the uniaxial in-plane anisotropy, which presumably originates from the film growth process. In an infinitely extended film without any anisotropies, the magnetization points always along the external magnetic field. In the present case, this implies that the magnetization always remains at 45° to the wave vector of the SAWs. For CoFeB rectangles of finite width, however, the anisotropy causes the magnetization not to point in the same direction as the external field at low magnetic fields, but to turn towards the direction of the long axis (x direction). This is clearly shown by the isofrequency curves in Fig. \ref{fig:fig3} a)-c), which are extracted from micromagnetic simulations using a spin-wave point source. The isofrequency curves for an infinitely extended film, which are shown by the dashed black lines, were simulated using periodic boundary conditions. These isofrequency curves are always rotated 45° to the wave vector of the SAWs. The isofrequency curves for finite structures, shown in blue, on the other hand, rotate with decreasing external field towards the propagation direction of the SAWs. In the present system this makes the overlap between the wave vector of the SAWs and that of the spin waves possible in the first place.\\
	
	To conclude, in this work, we have studied the excitation of spin waves by surface acoustic waves in micrometer-sized CoFeB rectangles. Our investigations with micro-focused BLS spectroscopy have demonstrated that spin waves can be excited lineally with the frequency of the SAW at two different magnetic fields. One excitation is identified as the FMR. The second excitation appears at low external magnetic fields. Through micromagnetic simulations, we were able to show that for this resonant excitation, the wave vector of the of the created spin wave is equal to the one of the exciting SAW. Thus,energy and linear momentum are conserved in this phonon-magnon scattering process. 
	
	\begin{acknowledgments}
		Financial support by the EU Horizon 2020 research and innovation program within the CHIRON project (contract no. 801055) and by the Deutsche Forschungsgemeinschaft (DFG, German Research Foundation) - TRR 173 "Spin+X"- 268565370 (Project B01) is gratefully acknowledged. DN acknowledges financial support from the Research Foundation - Flanders (FWO) through Grants 1SB9121N. We acknowledge valuable discussions with M. Weiler. 
	\end{acknowledgments}
	
	%
	%
	%
	\bibliographystyle{apsrev4-2}
	\bibliography{MSP-template}

\begin{thebibliography}{42}%
\makeatletter
\providecommand \@ifxundefined [1]{%
 \@ifx{#1\undefined}
}%
\providecommand \@ifnum [1]{%
 \ifnum #1\expandafter \@firstoftwo
 \else \expandafter \@secondoftwo
 \fi
}%
\providecommand \@ifx [1]{%
 \ifx #1\expandafter \@firstoftwo
 \else \expandafter \@secondoftwo
 \fi
}%
\providecommand \natexlab [1]{#1}%
\providecommand \enquote  [1]{``#1''}%
\providecommand \bibnamefont  [1]{#1}%
\providecommand \bibfnamefont [1]{#1}%
\providecommand \citenamefont [1]{#1}%
\providecommand \href@noop [0]{\@secondoftwo}%
\providecommand \href [0]{\begingroup \@sanitize@url \@href}%
\providecommand \@href[1]{\@@startlink{#1}\@@href}%
\providecommand \@@href[1]{\endgroup#1\@@endlink}%
\providecommand \@sanitize@url [0]{\catcode `\\12\catcode `\$12\catcode
  `\&12\catcode `\#12\catcode `\^12\catcode `\_12\catcode `\%12\relax}%
\providecommand \@@startlink[1]{}%
\providecommand \@@endlink[0]{}%
\providecommand \url  [0]{\begingroup\@sanitize@url \@url }%
\providecommand \@url [1]{\endgroup\@href {#1}{\urlprefix }}%
\providecommand \urlprefix  [0]{URL }%
\providecommand \Eprint [0]{\href }%
\providecommand \doibase [0]{https://doi.org/}%
\providecommand \selectlanguage [0]{\@gobble}%
\providecommand \bibinfo  [0]{\@secondoftwo}%
\providecommand \bibfield  [0]{\@secondoftwo}%
\providecommand \translation [1]{[#1]}%
\providecommand \BibitemOpen [0]{}%
\providecommand \bibitemStop [0]{}%
\providecommand \bibitemNoStop [0]{.\EOS\space}%
\providecommand \EOS [0]{\spacefactor3000\relax}%
\providecommand \BibitemShut  [1]{\csname bibitem#1\endcsname}%
\let\auto@bib@innerbib\@empty
\bibitem [{\citenamefont {Moore}(1965)}]{Moore1965}%
  \BibitemOpen
  \bibfield  {author} {\bibinfo {author} {\bibfnamefont {G.~E.}\ \bibnamefont
  {Moore}},\ }\href@noop {} {\bibfield  {journal} {\bibinfo  {journal}
  {Electronics}\ }\textbf {\bibinfo {volume} {114}} (\bibinfo {year}
  {1965})}\BibitemShut {NoStop}%
\bibitem [{\citenamefont {Serga}\ \emph {et~al.}(2010)\citenamefont {Serga},
  \citenamefont {Chumak},\ and\ \citenamefont {Hillebrands}}]{Serga2010}%
  \BibitemOpen
  \bibfield  {author} {\bibinfo {author} {\bibfnamefont {A.~A.}\ \bibnamefont
  {Serga}}, \bibinfo {author} {\bibfnamefont {A.~V.}\ \bibnamefont {Chumak}},\
  and\ \bibinfo {author} {\bibfnamefont {B.}~\bibnamefont {Hillebrands}},\
  }\bibfield  {journal} {\bibinfo  {journal} {Journal of Physics D: Applied
  Physics}\ }\textbf {\bibinfo {volume} {43}},\ \href
  {https://doi.org/10.1088/0022-3727/43/26/264002}
  {10.1088/0022-3727/43/26/264002} (\bibinfo {year} {2010})\BibitemShut
  {NoStop}%
\bibitem [{\citenamefont {Khitun}\ \emph {et~al.}(2010)\citenamefont {Khitun},
  \citenamefont {Bao},\ and\ \citenamefont {Wang}}]{Khitun2010}%
  \BibitemOpen
  \bibfield  {author} {\bibinfo {author} {\bibfnamefont {A.}~\bibnamefont
  {Khitun}}, \bibinfo {author} {\bibfnamefont {M.}~\bibnamefont {Bao}},\ and\
  \bibinfo {author} {\bibfnamefont {K.~L.}\ \bibnamefont {Wang}},\ }\bibfield
  {journal} {\bibinfo  {journal} {Journal of Physics D: Applied Physics}\
  }\textbf {\bibinfo {volume} {43}},\ \href
  {https://doi.org/10.1088/0022-3727/43/26/264005}
  {10.1088/0022-3727/43/26/264005} (\bibinfo {year} {2010})\BibitemShut
  {NoStop}%
\bibitem [{\citenamefont {Chumak}\ \emph {et~al.}(2015)\citenamefont {Chumak},
  \citenamefont {Vasyuchka}, \citenamefont {Serga},\ and\ \citenamefont
  {Hillebrands}}]{Chumak2015}%
  \BibitemOpen
  \bibfield  {author} {\bibinfo {author} {\bibfnamefont {A.~V.}\ \bibnamefont
  {Chumak}}, \bibinfo {author} {\bibfnamefont {V.~I.}\ \bibnamefont
  {Vasyuchka}}, \bibinfo {author} {\bibfnamefont {A.~A.}\ \bibnamefont
  {Serga}},\ and\ \bibinfo {author} {\bibfnamefont {B.}~\bibnamefont
  {Hillebrands}},\ }\href {https://doi.org/10.1038/nphys3347} {\bibfield
  {journal} {\bibinfo  {journal} {Nature Physics}\ }\textbf {\bibinfo {volume}
  {11}},\ \bibinfo {pages} {453} (\bibinfo {year} {2015})}\BibitemShut
  {NoStop}%
\bibitem [{\citenamefont {Kruglyak}\ \emph {et~al.}(2010)\citenamefont
  {Kruglyak}, \citenamefont {Demokritov},\ and\ \citenamefont
  {Grundler}}]{Kruglyak2010}%
  \BibitemOpen
  \bibfield  {author} {\bibinfo {author} {\bibfnamefont {V.}~\bibnamefont
  {Kruglyak}}, \bibinfo {author} {\bibfnamefont {O.}~\bibnamefont
  {Demokritov}},\ and\ \bibinfo {author} {\bibfnamefont {D.}~\bibnamefont
  {Grundler}},\ }\href {https://doi.org/10.1088/0022-3727/43/26/260301}
  {\bibfield  {journal} {\bibinfo  {journal} {Journal of Physics D: Applied
  Physics}\ }\textbf {\bibinfo {volume} {43}},\ \bibinfo {pages} {21} (\bibinfo
  {year} {2010})}\BibitemShut {NoStop}%
\bibitem [{\citenamefont {Pirro}\ \emph {et~al.}(2021)\citenamefont {Pirro},
  \citenamefont {Vasyuchka}, \citenamefont {Serga},\ and\ \citenamefont
  {Hillebrands}}]{Pirro2021}%
  \BibitemOpen
  \bibfield  {author} {\bibinfo {author} {\bibfnamefont {P.}~\bibnamefont
  {Pirro}}, \bibinfo {author} {\bibfnamefont {V.}~\bibnamefont {Vasyuchka}},
  \bibinfo {author} {\bibfnamefont {A.~A.}\ \bibnamefont {Serga}},\ and\
  \bibinfo {author} {\bibfnamefont {B.}~\bibnamefont {Hillebrands}},\ }\href
  {https://doi.org/10.1038/s41578-021-00332-w} {\bibfield  {journal} {\bibinfo
  {journal} {Nature Reviews Materials}\ }\textbf {\bibinfo {volume} {6}},\
  \bibinfo {pages} {1114} (\bibinfo {year} {2021})}\BibitemShut {NoStop}%
\bibitem [{\citenamefont {Papp}\ \emph {et~al.}(2021)\citenamefont {Papp},
  \citenamefont {Porod},\ and\ \citenamefont {Csaba}}]{Papp2021}%
  \BibitemOpen
  \bibfield  {author} {\bibinfo {author} {\bibfnamefont {A.}~\bibnamefont
  {Papp}}, \bibinfo {author} {\bibfnamefont {W.}~\bibnamefont {Porod}},\ and\
  \bibinfo {author} {\bibfnamefont {G.}~\bibnamefont {Csaba}},\ }\href@noop {}
  {\bibfield  {journal} {\bibinfo  {journal} {Nature Communications}\ }\textbf
  {\bibinfo {volume} {12}},\ \bibinfo {pages} {6422} (\bibinfo {year}
  {2021})}\BibitemShut {NoStop}%
\bibitem [{\citenamefont {Chumak}\ \emph {et~al.}(2014)\citenamefont {Chumak},
  \citenamefont {Serga},\ and\ \citenamefont {Hillebrands}}]{Chumak2014}%
  \BibitemOpen
  \bibfield  {author} {\bibinfo {author} {\bibfnamefont {A.~V.}\ \bibnamefont
  {Chumak}}, \bibinfo {author} {\bibfnamefont {A.~A.}\ \bibnamefont {Serga}},\
  and\ \bibinfo {author} {\bibfnamefont {B.}~\bibnamefont {Hillebrands}},\
  }\href {https://doi.org/10.1038/ncomms5700} {\bibfield  {journal} {\bibinfo
  {journal} {Nature Communications}\ }\textbf {\bibinfo {volume} {5}},\
  \bibinfo {pages} {1} (\bibinfo {year} {2014})}\BibitemShut {NoStop}%
\bibitem [{\citenamefont {Br{\"{a}}cher}\ \emph {et~al.}(2017)\citenamefont
  {Br{\"{a}}cher}, \citenamefont {Boulle}, \citenamefont {Gaudin},\ and\
  \citenamefont {Pirro}}]{Bracher2017}%
  \BibitemOpen
  \bibfield  {author} {\bibinfo {author} {\bibfnamefont {T.}~\bibnamefont
  {Br{\"{a}}cher}}, \bibinfo {author} {\bibfnamefont {O.}~\bibnamefont
  {Boulle}}, \bibinfo {author} {\bibfnamefont {G.}~\bibnamefont {Gaudin}},\
  and\ \bibinfo {author} {\bibfnamefont {P.}~\bibnamefont {Pirro}},\ }\href
  {https://doi.org/10.1103/PhysRevB.95.064429} {\bibfield  {journal} {\bibinfo
  {journal} {Physical Review B}\ }\textbf {\bibinfo {volume} {95}},\ \bibinfo
  {pages} {1} (\bibinfo {year} {2017})},\ \Eprint
  {https://arxiv.org/abs/1611.07841} {arXiv:1611.07841} \BibitemShut {NoStop}%
\bibitem [{\citenamefont {Grassi}\ \emph {et~al.}(2020)\citenamefont {Grassi},
  \citenamefont {Geilen}, \citenamefont {Louis}, \citenamefont {Mohseni},
  \citenamefont {Br{\"{a}}cher}, \citenamefont {Hehn}, \citenamefont
  {Stoeffler}, \citenamefont {Bailleul}, \citenamefont {Pirro},\ and\
  \citenamefont {Henry}}]{Grassi2020}%
  \BibitemOpen
  \bibfield  {author} {\bibinfo {author} {\bibfnamefont {M.}~\bibnamefont
  {Grassi}}, \bibinfo {author} {\bibfnamefont {M.}~\bibnamefont {Geilen}},
  \bibinfo {author} {\bibfnamefont {D.}~\bibnamefont {Louis}}, \bibinfo
  {author} {\bibfnamefont {M.}~\bibnamefont {Mohseni}}, \bibinfo {author}
  {\bibfnamefont {T.}~\bibnamefont {Br{\"{a}}cher}}, \bibinfo {author}
  {\bibfnamefont {M.}~\bibnamefont {Hehn}}, \bibinfo {author} {\bibfnamefont
  {D.}~\bibnamefont {Stoeffler}}, \bibinfo {author} {\bibfnamefont
  {M.}~\bibnamefont {Bailleul}}, \bibinfo {author} {\bibfnamefont
  {P.}~\bibnamefont {Pirro}},\ and\ \bibinfo {author} {\bibfnamefont
  {Y.}~\bibnamefont {Henry}},\ }\href
  {https://doi.org/10.1103/PhysRevApplied.14.024047} {\bibfield  {journal}
  {\bibinfo  {journal} {Physical Review Applied}\ }\textbf {\bibinfo {volume}
  {14}},\ \bibinfo {pages} {1} (\bibinfo {year} {2020})}\BibitemShut {NoStop}%
\bibitem [{\citenamefont {Szulc}\ \emph {et~al.}(2020)\citenamefont {Szulc},
  \citenamefont {Graczyk}, \citenamefont {Mruczkiewicz}, \citenamefont
  {Gubbiotti},\ and\ \citenamefont {Krawczyk}}]{Szulc2020}%
  \BibitemOpen
  \bibfield  {author} {\bibinfo {author} {\bibfnamefont {K.}~\bibnamefont
  {Szulc}}, \bibinfo {author} {\bibfnamefont {P.}~\bibnamefont {Graczyk}},
  \bibinfo {author} {\bibfnamefont {M.}~\bibnamefont {Mruczkiewicz}}, \bibinfo
  {author} {\bibfnamefont {G.}~\bibnamefont {Gubbiotti}},\ and\ \bibinfo
  {author} {\bibfnamefont {M.}~\bibnamefont {Krawczyk}},\ }\href
  {https://doi.org/10.1103/PhysRevApplied.14.034063} {\bibfield  {journal}
  {\bibinfo  {journal} {Physical Review Applied}\ }\textbf {\bibinfo {volume}
  {14}},\ \bibinfo {pages} {1} (\bibinfo {year} {2020})},\ \Eprint
  {https://arxiv.org/abs/2002.06096} {arXiv:2002.06096} \BibitemShut {NoStop}%
\bibitem [{\citenamefont {Klingler}\ \emph {et~al.}(2014)\citenamefont
  {Klingler}, \citenamefont {Pirro}, \citenamefont {Br{\"{a}}cher},
  \citenamefont {Leven}, \citenamefont {Hillebrands},\ and\ \citenamefont
  {Chumak}}]{Klingler2014}%
  \BibitemOpen
  \bibfield  {author} {\bibinfo {author} {\bibfnamefont {S.}~\bibnamefont
  {Klingler}}, \bibinfo {author} {\bibfnamefont {P.}~\bibnamefont {Pirro}},
  \bibinfo {author} {\bibfnamefont {T.}~\bibnamefont {Br{\"{a}}cher}}, \bibinfo
  {author} {\bibfnamefont {B.}~\bibnamefont {Leven}}, \bibinfo {author}
  {\bibfnamefont {B.}~\bibnamefont {Hillebrands}},\ and\ \bibinfo {author}
  {\bibfnamefont {A.~V.}\ \bibnamefont {Chumak}},\ }\href
  {https://doi.org/10.1063/1.4898042} {\bibfield  {journal} {\bibinfo
  {journal} {Applied Physics Letters}\ }\textbf {\bibinfo {volume} {105}},\
  \bibinfo {pages} {1} (\bibinfo {year} {2014})}\BibitemShut {NoStop}%
\bibitem [{\citenamefont {Fischer}\ \emph {et~al.}(2017)\citenamefont
  {Fischer}, \citenamefont {Kewenig}, \citenamefont {Bozhko}, \citenamefont
  {Serga}, \citenamefont {Syvorotka}, \citenamefont {Ciubotaru}, \citenamefont
  {Adelmann}, \citenamefont {Hillebrands},\ and\ \citenamefont
  {Chumak}}]{Fischer2017}%
  \BibitemOpen
  \bibfield  {author} {\bibinfo {author} {\bibfnamefont {T.}~\bibnamefont
  {Fischer}}, \bibinfo {author} {\bibfnamefont {M.}~\bibnamefont {Kewenig}},
  \bibinfo {author} {\bibfnamefont {D.~A.}\ \bibnamefont {Bozhko}}, \bibinfo
  {author} {\bibfnamefont {A.~A.}\ \bibnamefont {Serga}}, \bibinfo {author}
  {\bibfnamefont {I.~I.}\ \bibnamefont {Syvorotka}}, \bibinfo {author}
  {\bibfnamefont {F.}~\bibnamefont {Ciubotaru}}, \bibinfo {author}
  {\bibfnamefont {C.}~\bibnamefont {Adelmann}}, \bibinfo {author}
  {\bibfnamefont {B.}~\bibnamefont {Hillebrands}},\ and\ \bibinfo {author}
  {\bibfnamefont {A.~V.}\ \bibnamefont {Chumak}},\ }\bibfield  {journal}
  {\bibinfo  {journal} {Applied Physics Letters}\ }\textbf {\bibinfo {volume}
  {110}},\ \href {https://doi.org/10.1063/1.4979840} {10.1063/1.4979840}
  (\bibinfo {year} {2017}),\ \Eprint {https://arxiv.org/abs/1612.07708}
  {arXiv:1612.07708} \BibitemShut {NoStop}%
\bibitem [{\citenamefont {Talmelli}\ \emph {et~al.}(2020)\citenamefont
  {Talmelli}, \citenamefont {Devolder}, \citenamefont {Tr{\"{a}}ger},
  \citenamefont {F{\"{o}}rster}, \citenamefont {Wintz}, \citenamefont
  {Weigand}, \citenamefont {Stoll}, \citenamefont {Heyns}, \citenamefont
  {Sch{\"{u}}tz}, \citenamefont {Radu}, \citenamefont {Gr{\"{a}}fe},
  \citenamefont {Ciubotaru},\ and\ \citenamefont {Adelmann}}]{Talmelli2020}%
  \BibitemOpen
  \bibfield  {author} {\bibinfo {author} {\bibfnamefont {G.}~\bibnamefont
  {Talmelli}}, \bibinfo {author} {\bibfnamefont {T.}~\bibnamefont {Devolder}},
  \bibinfo {author} {\bibfnamefont {N.}~\bibnamefont {Tr{\"{a}}ger}}, \bibinfo
  {author} {\bibfnamefont {J.}~\bibnamefont {F{\"{o}}rster}}, \bibinfo {author}
  {\bibfnamefont {S.}~\bibnamefont {Wintz}}, \bibinfo {author} {\bibfnamefont
  {M.}~\bibnamefont {Weigand}}, \bibinfo {author} {\bibfnamefont
  {H.}~\bibnamefont {Stoll}}, \bibinfo {author} {\bibfnamefont
  {M.}~\bibnamefont {Heyns}}, \bibinfo {author} {\bibfnamefont
  {G.}~\bibnamefont {Sch{\"{u}}tz}}, \bibinfo {author} {\bibfnamefont {I.~P.}\
  \bibnamefont {Radu}}, \bibinfo {author} {\bibfnamefont {J.}~\bibnamefont
  {Gr{\"{a}}fe}}, \bibinfo {author} {\bibfnamefont {F.}~\bibnamefont
  {Ciubotaru}},\ and\ \bibinfo {author} {\bibfnamefont {C.}~\bibnamefont
  {Adelmann}},\ }\href
  {https://doi.org/10.1126/SCIADV.ABB4042/SUPPL_FILE/ABB4042_SM.PDF} {\bibfield
   {journal} {\bibinfo  {journal} {Science Advances}\ }\textbf {\bibinfo
  {volume} {6}},\ \bibinfo {pages} {4042} (\bibinfo {year} {2020})}\BibitemShut
  {NoStop}%
\bibitem [{\citenamefont {Sadovnikov}\ \emph {et~al.}(2015)\citenamefont
  {Sadovnikov}, \citenamefont {Beginin}, \citenamefont {Sheshukova},
  \citenamefont {Romanenko}, \citenamefont {Sharaevskii},\ and\ \citenamefont
  {Nikitov}}]{Sadovnikov2015}%
  \BibitemOpen
  \bibfield  {author} {\bibinfo {author} {\bibfnamefont {A.~V.}\ \bibnamefont
  {Sadovnikov}}, \bibinfo {author} {\bibfnamefont {E.~N.}\ \bibnamefont
  {Beginin}}, \bibinfo {author} {\bibfnamefont {S.~E.}\ \bibnamefont
  {Sheshukova}}, \bibinfo {author} {\bibfnamefont {D.~V.}\ \bibnamefont
  {Romanenko}}, \bibinfo {author} {\bibfnamefont {Y.~P.}\ \bibnamefont
  {Sharaevskii}},\ and\ \bibinfo {author} {\bibfnamefont {S.~A.}\ \bibnamefont
  {Nikitov}},\ }\bibfield  {journal} {\bibinfo  {journal} {Applied Physics
  Letters}\ }\textbf {\bibinfo {volume} {107}},\ \href
  {https://doi.org/10.1063/1.4936207} {10.1063/1.4936207} (\bibinfo {year}
  {2015})\BibitemShut {NoStop}%
\bibitem [{\citenamefont {Wang}\ \emph {et~al.}(2020)\citenamefont {Wang},
  \citenamefont {Kewenig}, \citenamefont {Schneider}, \citenamefont {Verba},
  \citenamefont {Kohl}, \citenamefont {Heinz}, \citenamefont {Geilen},
  \citenamefont {Mohseni}, \citenamefont {Lägel}, \citenamefont {Ciubotaru},
  \citenamefont {Adelmann}, \citenamefont {Dubs}, \citenamefont {Cotofana},
  \citenamefont {Dobrovolskiy}, \citenamefont {Brächer}, \citenamefont
  {Pirro},\ and\ \citenamefont {Chumak}}]{Qi2020}%
  \BibitemOpen
  \bibfield  {author} {\bibinfo {author} {\bibfnamefont {Q.}~\bibnamefont
  {Wang}}, \bibinfo {author} {\bibfnamefont {M.}~\bibnamefont {Kewenig}},
  \bibinfo {author} {\bibfnamefont {M.}~\bibnamefont {Schneider}}, \bibinfo
  {author} {\bibfnamefont {R.}~\bibnamefont {Verba}}, \bibinfo {author}
  {\bibfnamefont {F.}~\bibnamefont {Kohl}}, \bibinfo {author} {\bibfnamefont
  {B.}~\bibnamefont {Heinz}}, \bibinfo {author} {\bibfnamefont
  {M.}~\bibnamefont {Geilen}}, \bibinfo {author} {\bibfnamefont
  {M.}~\bibnamefont {Mohseni}}, \bibinfo {author} {\bibfnamefont
  {B.}~\bibnamefont {Lägel}}, \bibinfo {author} {\bibfnamefont
  {F.}~\bibnamefont {Ciubotaru}}, \bibinfo {author} {\bibfnamefont
  {C.}~\bibnamefont {Adelmann}}, \bibinfo {author} {\bibfnamefont
  {C.}~\bibnamefont {Dubs}}, \bibinfo {author} {\bibfnamefont {S.~D.}\
  \bibnamefont {Cotofana}}, \bibinfo {author} {\bibfnamefont {O.~V.}\
  \bibnamefont {Dobrovolskiy}}, \bibinfo {author} {\bibfnamefont
  {T.}~\bibnamefont {Brächer}}, \bibinfo {author} {\bibfnamefont
  {P.}~\bibnamefont {Pirro}},\ and\ \bibinfo {author} {\bibfnamefont {A.~V.}\
  \bibnamefont {Chumak}},\ }\href@noop {} {\bibfield  {journal} {\bibinfo
  {journal} {Nature electronics}\ }\textbf {\bibinfo {volume} {3}} (\bibinfo
  {year} {2020})}\BibitemShut {NoStop}%
\bibitem [{\citenamefont {Mahmoud}\ \emph {et~al.}(2020)\citenamefont
  {Mahmoud}, \citenamefont {Ciubotaru}, \citenamefont {Vanderveken},
  \citenamefont {Chumak}, \citenamefont {Hamdioui}, \citenamefont {Adelmann},\
  and\ \citenamefont {Cotofana}}]{Mahmoud2020}%
  \BibitemOpen
  \bibfield  {author} {\bibinfo {author} {\bibfnamefont {A.}~\bibnamefont
  {Mahmoud}}, \bibinfo {author} {\bibfnamefont {F.}~\bibnamefont {Ciubotaru}},
  \bibinfo {author} {\bibfnamefont {F.}~\bibnamefont {Vanderveken}}, \bibinfo
  {author} {\bibfnamefont {A.~V.}\ \bibnamefont {Chumak}}, \bibinfo {author}
  {\bibfnamefont {S.}~\bibnamefont {Hamdioui}}, \bibinfo {author}
  {\bibfnamefont {C.}~\bibnamefont {Adelmann}},\ and\ \bibinfo {author}
  {\bibfnamefont {S.}~\bibnamefont {Cotofana}},\ }\bibfield  {journal}
  {\bibinfo  {journal} {Journal of Applied Physics}\ }\textbf {\bibinfo
  {volume} {128}},\ \href {https://doi.org/10.1063/5.0019328}
  {10.1063/5.0019328} (\bibinfo {year} {2020}),\ \Eprint
  {https://arxiv.org/abs/2006.12905} {arXiv:2006.12905} \BibitemShut {NoStop}%
\bibitem [{\citenamefont {Fiebig}(2005)}]{Fiebig2005}%
  \BibitemOpen
  \bibfield  {author} {\bibinfo {author} {\bibfnamefont {M.}~\bibnamefont
  {Fiebig}},\ }\bibfield  {journal} {\bibinfo  {journal} {Journal of Physics D:
  Applied Physics}\ }\textbf {\bibinfo {volume} {38}},\ \href
  {https://doi.org/10.1088/0022-3727/38/8/R01} {10.1088/0022-3727/38/8/R01}
  (\bibinfo {year} {2005})\BibitemShut {NoStop}%
\bibitem [{\citenamefont {Smolenskiĭ}\ and\ \citenamefont
  {Chupis}(1982)}]{Smolenskii1982}%
  \BibitemOpen
  \bibfield  {author} {\bibinfo {author} {\bibfnamefont {G.~A.}\ \bibnamefont
  {Smolenskiĭ}}\ and\ \bibinfo {author} {\bibfnamefont {I.~E.}\ \bibnamefont
  {Chupis}},\ }\href {https://doi.org/10.1070/PU1982v025n07ABEH004570}
  {\bibfield  {journal} {\bibinfo  {journal} {Soviet Physics Uspekhi}\ }\textbf
  {\bibinfo {volume} {25}},\ \bibinfo {pages} {475} (\bibinfo {year}
  {1982})}\BibitemShut {NoStop}%
\bibitem [{\citenamefont {Shelukhin}\ \emph {et~al.}(2020)\citenamefont
  {Shelukhin}, \citenamefont {Pertsev}, \citenamefont {Scherbakov},
  \citenamefont {Kazenwadel}, \citenamefont {Kirilenko}, \citenamefont
  {H{\"{a}}m{\"{a}}la{\"{i}}nen}, \citenamefont {{Van Dijken}},\ and\
  \citenamefont {Kalashnikova}}]{Shelukhin2020}%
  \BibitemOpen
  \bibfield  {author} {\bibinfo {author} {\bibfnamefont {L.~A.}\ \bibnamefont
  {Shelukhin}}, \bibinfo {author} {\bibfnamefont {N.~A.}\ \bibnamefont
  {Pertsev}}, \bibinfo {author} {\bibfnamefont {A.~V.}\ \bibnamefont
  {Scherbakov}}, \bibinfo {author} {\bibfnamefont {D.~L.}\ \bibnamefont
  {Kazenwadel}}, \bibinfo {author} {\bibfnamefont {D.~A.}\ \bibnamefont
  {Kirilenko}}, \bibinfo {author} {\bibfnamefont {S.~J.}\ \bibnamefont
  {H{\"{a}}m{\"{a}}la{\"{i}}nen}}, \bibinfo {author} {\bibfnamefont
  {S.}~\bibnamefont {{Van Dijken}}},\ and\ \bibinfo {author} {\bibfnamefont
  {A.~M.}\ \bibnamefont {Kalashnikova}},\ }\href
  {https://doi.org/10.1103/PhysRevApplied.14.034061} {\bibfield  {journal}
  {\bibinfo  {journal} {Physical Review Applied}\ }\textbf {\bibinfo {volume}
  {14}},\ \bibinfo {pages} {1} (\bibinfo {year} {2020})},\ \Eprint
  {https://arxiv.org/abs/2004.03566} {arXiv:2004.03566} \BibitemShut {NoStop}%
\bibitem [{\citenamefont {Khitun}\ and\ \citenamefont
  {Wang}(2011)}]{Khitun2011a}%
  \BibitemOpen
  \bibfield  {author} {\bibinfo {author} {\bibfnamefont {A.}~\bibnamefont
  {Khitun}}\ and\ \bibinfo {author} {\bibfnamefont {K.~L.}\ \bibnamefont
  {Wang}},\ }\href {https://doi.org/10.1063/1.3609062} {\bibfield  {journal}
  {\bibinfo  {journal} {Journal of Applied Physics}\ }\textbf {\bibinfo
  {volume} {110}},\ \bibinfo {pages} {034306} (\bibinfo {year}
  {2011})}\BibitemShut {NoStop}%
\bibitem [{\citenamefont {Alekseev}\ \emph {et~al.}(2020)\citenamefont
  {Alekseev}, \citenamefont {Dizhur}, \citenamefont {Polzikova}, \citenamefont
  {Luzanov}, \citenamefont {Raevskiy}, \citenamefont {Orlov}, \citenamefont
  {Kotov},\ and\ \citenamefont {Nikitov}}]{Alekseev2020}%
  \BibitemOpen
  \bibfield  {author} {\bibinfo {author} {\bibfnamefont {S.~G.}\ \bibnamefont
  {Alekseev}}, \bibinfo {author} {\bibfnamefont {S.~E.}\ \bibnamefont
  {Dizhur}}, \bibinfo {author} {\bibfnamefont {N.~I.}\ \bibnamefont
  {Polzikova}}, \bibinfo {author} {\bibfnamefont {V.~A.}\ \bibnamefont
  {Luzanov}}, \bibinfo {author} {\bibfnamefont {A.~O.}\ \bibnamefont
  {Raevskiy}}, \bibinfo {author} {\bibfnamefont {A.~P.}\ \bibnamefont {Orlov}},
  \bibinfo {author} {\bibfnamefont {V.~A.}\ \bibnamefont {Kotov}},\ and\
  \bibinfo {author} {\bibfnamefont {S.~A.}\ \bibnamefont {Nikitov}},\ }\href
  {https://doi.org/10.1063/5.0022267} {\bibfield  {journal} {\bibinfo
  {journal} {Applied Physics Letters}\ }\textbf {\bibinfo {volume} {117}},\
  \bibinfo {pages} {20} (\bibinfo {year} {2020})},\ \Eprint
  {https://arxiv.org/abs/2008.09520} {arXiv:2008.09520} \BibitemShut {NoStop}%
\bibitem [{\citenamefont {Dreher}\ \emph {et~al.}(2012)\citenamefont {Dreher},
  \citenamefont {Weiler}, \citenamefont {Pernpeintner}, \citenamefont {Huebl},
  \citenamefont {Gross}, \citenamefont {Brandt},\ and\ \citenamefont
  {Goennenwein}}]{Dreher2012}%
  \BibitemOpen
  \bibfield  {author} {\bibinfo {author} {\bibfnamefont {L.}~\bibnamefont
  {Dreher}}, \bibinfo {author} {\bibfnamefont {M.}~\bibnamefont {Weiler}},
  \bibinfo {author} {\bibfnamefont {M.}~\bibnamefont {Pernpeintner}}, \bibinfo
  {author} {\bibfnamefont {H.}~\bibnamefont {Huebl}}, \bibinfo {author}
  {\bibfnamefont {R.}~\bibnamefont {Gross}}, \bibinfo {author} {\bibfnamefont
  {M.~S.}\ \bibnamefont {Brandt}},\ and\ \bibinfo {author} {\bibfnamefont
  {S.~T.}\ \bibnamefont {Goennenwein}},\ }\href
  {https://doi.org/10.1103/PhysRevB.86.134415} {\bibfield  {journal} {\bibinfo
  {journal} {Physical Review B - Condensed Matter and Materials Physics}\
  }\textbf {\bibinfo {volume} {86}},\ \bibinfo {pages} {1} (\bibinfo {year}
  {2012})},\ \Eprint {https://arxiv.org/abs/1208.0001} {arXiv:1208.0001}
  \BibitemShut {NoStop}%
\bibitem [{\citenamefont {K{\"{u}}{\ss}}\ \emph {et~al.}(2021)\citenamefont
  {K{\"{u}}{\ss}}, \citenamefont {Heigl}, \citenamefont {Flacke}, \citenamefont
  {Hefele}, \citenamefont {H{\"{o}}rner}, \citenamefont {Weiler}, \citenamefont
  {Albrecht},\ and\ \citenamefont {Wixforth}}]{Kuss2021}%
  \BibitemOpen
  \bibfield  {author} {\bibinfo {author} {\bibfnamefont {M.}~\bibnamefont
  {K{\"{u}}{\ss}}}, \bibinfo {author} {\bibfnamefont {M.}~\bibnamefont
  {Heigl}}, \bibinfo {author} {\bibfnamefont {L.}~\bibnamefont {Flacke}},
  \bibinfo {author} {\bibfnamefont {A.}~\bibnamefont {Hefele}}, \bibinfo
  {author} {\bibfnamefont {A.}~\bibnamefont {H{\"{o}}rner}}, \bibinfo {author}
  {\bibfnamefont {M.}~\bibnamefont {Weiler}}, \bibinfo {author} {\bibfnamefont
  {M.}~\bibnamefont {Albrecht}},\ and\ \bibinfo {author} {\bibfnamefont
  {A.}~\bibnamefont {Wixforth}},\ }\href
  {https://doi.org/10.1103/PhysRevApplied.15.034046} {\bibfield  {journal}
  {\bibinfo  {journal} {Physical Review Applied}\ }\textbf {\bibinfo {volume}
  {15}},\ \bibinfo {pages} {034046} (\bibinfo {year} {2021})},\ \Eprint
  {https://arxiv.org/abs/2012.01055} {arXiv:2012.01055} \BibitemShut {NoStop}%
\bibitem [{\citenamefont {Verba}\ \emph {et~al.}(2018)\citenamefont {Verba},
  \citenamefont {Lisenkov}, \citenamefont {Krivorotov}, \citenamefont
  {Tiberkevich},\ and\ \citenamefont {Slavin}}]{Verba2018}%
  \BibitemOpen
  \bibfield  {author} {\bibinfo {author} {\bibfnamefont {R.}~\bibnamefont
  {Verba}}, \bibinfo {author} {\bibfnamefont {I.}~\bibnamefont {Lisenkov}},
  \bibinfo {author} {\bibfnamefont {I.}~\bibnamefont {Krivorotov}}, \bibinfo
  {author} {\bibfnamefont {V.}~\bibnamefont {Tiberkevich}},\ and\ \bibinfo
  {author} {\bibfnamefont {A.}~\bibnamefont {Slavin}},\ }\href
  {https://doi.org/10.1103/PhysRevApplied.9.064014} {\bibfield  {journal}
  {\bibinfo  {journal} {Physical Review Applied}\ }\textbf {\bibinfo {volume}
  {9}},\ \bibinfo {pages} {64014} (\bibinfo {year} {2018})}\BibitemShut
  {NoStop}%
\bibitem [{\citenamefont {K{\"{u}}\ss}\ \emph {et~al.}(2020)\citenamefont
  {K{\"{u}}\ss}, \citenamefont {Heigl}, \citenamefont {Flacke}, \citenamefont
  {H{\"{o}}rner}, \citenamefont {Weiler}, \citenamefont {Albrecht},\ and\
  \citenamefont {Wixforth}}]{Ku2020}%
  \BibitemOpen
  \bibfield  {author} {\bibinfo {author} {\bibfnamefont {M.}~\bibnamefont
  {K{\"{u}}\ss}}, \bibinfo {author} {\bibfnamefont {M.}~\bibnamefont {Heigl}},
  \bibinfo {author} {\bibfnamefont {L.}~\bibnamefont {Flacke}}, \bibinfo
  {author} {\bibfnamefont {A.}~\bibnamefont {H{\"{o}}rner}}, \bibinfo {author}
  {\bibfnamefont {M.}~\bibnamefont {Weiler}}, \bibinfo {author} {\bibfnamefont
  {M.}~\bibnamefont {Albrecht}},\ and\ \bibinfo {author} {\bibfnamefont
  {A.}~\bibnamefont {Wixforth}},\ }\href
  {https://doi.org/10.1103/PhysRevLett.125.217203} {\bibfield  {journal}
  {\bibinfo  {journal} {Physical Review Letters}\ }\textbf {\bibinfo {volume}
  {125}},\ \bibinfo {pages} {217203} (\bibinfo {year} {2020})},\ \Eprint
  {https://arxiv.org/abs/2004.03535} {arXiv:2004.03535} \BibitemShut {NoStop}%
\bibitem [{\citenamefont {Xu}\ \emph {et~al.}(2020)\citenamefont {Xu},
  \citenamefont {Yamamoto}, \citenamefont {Puebla}, \citenamefont {Baumgaertl},
  \citenamefont {Rana}, \citenamefont {Miura}, \citenamefont {Takahashi},
  \citenamefont {Grundler}, \citenamefont {Maekawa},\ and\ \citenamefont
  {Otani}}]{Xu2020}%
  \BibitemOpen
  \bibfield  {author} {\bibinfo {author} {\bibfnamefont {M.}~\bibnamefont
  {Xu}}, \bibinfo {author} {\bibfnamefont {K.}~\bibnamefont {Yamamoto}},
  \bibinfo {author} {\bibfnamefont {J.}~\bibnamefont {Puebla}}, \bibinfo
  {author} {\bibfnamefont {K.}~\bibnamefont {Baumgaertl}}, \bibinfo {author}
  {\bibfnamefont {B.}~\bibnamefont {Rana}}, \bibinfo {author} {\bibfnamefont
  {K.}~\bibnamefont {Miura}}, \bibinfo {author} {\bibfnamefont
  {H.}~\bibnamefont {Takahashi}}, \bibinfo {author} {\bibfnamefont
  {D.}~\bibnamefont {Grundler}}, \bibinfo {author} {\bibfnamefont
  {S.}~\bibnamefont {Maekawa}},\ and\ \bibinfo {author} {\bibfnamefont
  {Y.}~\bibnamefont {Otani}},\ }\href {https://doi.org/10.1126/sciadv.abb1724}
  {\bibfield  {journal} {\bibinfo  {journal} {Science Advances}\ }\textbf
  {\bibinfo {volume} {6}},\ \bibinfo {pages} {eabb1724} (\bibinfo {year}
  {2020})}\BibitemShut {NoStop}%
\bibitem [{\citenamefont {Babu}\ \emph {et~al.}(2020)\citenamefont {Babu},
  \citenamefont {Trzaskowska}, \citenamefont {Graczyk}, \citenamefont
  {Centa{\l}a}, \citenamefont {Mieszczak}, \citenamefont {G{\l}owi{\'{n}}ski},
  \citenamefont {Zdunek}, \citenamefont {Mielcarek},\ and\ \citenamefont
  {K{\l}os}}]{Babu2020}%
  \BibitemOpen
  \bibfield  {author} {\bibinfo {author} {\bibfnamefont {N.~K.}\ \bibnamefont
  {Babu}}, \bibinfo {author} {\bibfnamefont {A.}~\bibnamefont {Trzaskowska}},
  \bibinfo {author} {\bibfnamefont {P.}~\bibnamefont {Graczyk}}, \bibinfo
  {author} {\bibfnamefont {G.}~\bibnamefont {Centa{\l}a}}, \bibinfo {author}
  {\bibfnamefont {S.}~\bibnamefont {Mieszczak}}, \bibinfo {author}
  {\bibfnamefont {H.}~\bibnamefont {G{\l}owi{\'{n}}ski}}, \bibinfo {author}
  {\bibfnamefont {M.}~\bibnamefont {Zdunek}}, \bibinfo {author} {\bibfnamefont
  {S.}~\bibnamefont {Mielcarek}},\ and\ \bibinfo {author} {\bibfnamefont
  {J.~W.}\ \bibnamefont {K{\l}os}},\ }\bibfield  {journal} {\bibinfo  {journal}
  {Nano Letters}\ }\href {https://doi.org/10.1021/acs.nanolett.0c03692}
  {10.1021/acs.nanolett.0c03692} (\bibinfo {year} {2020})\BibitemShut {NoStop}%
\bibitem [{\citenamefont {Zhao}\ \emph {et~al.}(2021)\citenamefont {Zhao},
  \citenamefont {Zhang}, \citenamefont {Li}, \citenamefont {Zhang},
  \citenamefont {Pearson}, \citenamefont {Divan}, \citenamefont {Liu},
  \citenamefont {Novosad}, \citenamefont {Wang},\ and\ \citenamefont
  {Hoffmann}}]{Zhao2021}%
  \BibitemOpen
  \bibfield  {author} {\bibinfo {author} {\bibfnamefont {C.}~\bibnamefont
  {Zhao}}, \bibinfo {author} {\bibfnamefont {Z.}~\bibnamefont {Zhang}},
  \bibinfo {author} {\bibfnamefont {Y.}~\bibnamefont {Li}}, \bibinfo {author}
  {\bibfnamefont {W.}~\bibnamefont {Zhang}}, \bibinfo {author} {\bibfnamefont
  {J.~E.}\ \bibnamefont {Pearson}}, \bibinfo {author} {\bibfnamefont
  {R.}~\bibnamefont {Divan}}, \bibinfo {author} {\bibfnamefont
  {Q.}~\bibnamefont {Liu}}, \bibinfo {author} {\bibfnamefont {V.}~\bibnamefont
  {Novosad}}, \bibinfo {author} {\bibfnamefont {J.}~\bibnamefont {Wang}},\ and\
  \bibinfo {author} {\bibfnamefont {A.}~\bibnamefont {Hoffmann}},\ }\href
  {https://doi.org/10.1103/PhysRevApplied.15.014052} {\bibfield  {journal}
  {\bibinfo  {journal} {Physical Review Applied}\ }\textbf {\bibinfo {volume}
  {15}},\ \bibinfo {pages} {014052} (\bibinfo {year} {2021})},\ \Eprint
  {https://arxiv.org/abs/2008.11391} {arXiv:2008.11391} \BibitemShut {NoStop}%
\bibitem [{\citenamefont {Kraimia}\ \emph {et~al.}(2020)\citenamefont
  {Kraimia}, \citenamefont {Kuszewski}, \citenamefont {Duquesne}, \citenamefont
  {Lema{\^{i}}tre}, \citenamefont {Margaillan}, \citenamefont {Gourdon},\ and\
  \citenamefont {Thevenard}}]{Kraimia2020}%
  \BibitemOpen
  \bibfield  {author} {\bibinfo {author} {\bibfnamefont {M.}~\bibnamefont
  {Kraimia}}, \bibinfo {author} {\bibfnamefont {P.}~\bibnamefont {Kuszewski}},
  \bibinfo {author} {\bibfnamefont {J.~Y.}\ \bibnamefont {Duquesne}}, \bibinfo
  {author} {\bibfnamefont {A.}~\bibnamefont {Lema{\^{i}}tre}}, \bibinfo
  {author} {\bibfnamefont {F.}~\bibnamefont {Margaillan}}, \bibinfo {author}
  {\bibfnamefont {C.}~\bibnamefont {Gourdon}},\ and\ \bibinfo {author}
  {\bibfnamefont {L.}~\bibnamefont {Thevenard}},\ }\href
  {https://doi.org/10.1103/PhysRevB.101.144425} {\bibfield  {journal} {\bibinfo
   {journal} {Physical Review B}\ }\textbf {\bibinfo {volume} {101}},\ \bibinfo
  {pages} {144425} (\bibinfo {year} {2020})}\BibitemShut {NoStop}%
\bibitem [{\citenamefont {Sebastian}\ \emph {et~al.}(2015)\citenamefont
  {Sebastian}, \citenamefont {Schultheiss}, \citenamefont {Obry}, \citenamefont
  {Hillebrands},\ and\ \citenamefont {Schultheiss}}]{Sebastian2015}%
  \BibitemOpen
  \bibfield  {author} {\bibinfo {author} {\bibfnamefont {T.}~\bibnamefont
  {Sebastian}}, \bibinfo {author} {\bibfnamefont {K.}~\bibnamefont
  {Schultheiss}}, \bibinfo {author} {\bibfnamefont {B.}~\bibnamefont {Obry}},
  \bibinfo {author} {\bibfnamefont {B.}~\bibnamefont {Hillebrands}},\ and\
  \bibinfo {author} {\bibfnamefont {H.}~\bibnamefont {Schultheiss}},\ }\href
  {https://doi.org/10.3389/fphy.2015.00035} {\bibfield  {journal} {\bibinfo
  {journal} {Frontiers in Physics}\ }\textbf {\bibinfo {volume} {3}},\ \bibinfo
  {pages} {1} (\bibinfo {year} {2015})}\BibitemShut {NoStop}%
\bibitem [{\citenamefont {Geilen}\ \emph {et~al.}(2020)\citenamefont {Geilen},
  \citenamefont {Kohl}, \citenamefont {Nicoloiu}, \citenamefont {M{\"{u}}ller},
  \citenamefont {Hillebrands},\ and\ \citenamefont {Pirro}}]{Geilen2020}%
  \BibitemOpen
  \bibfield  {author} {\bibinfo {author} {\bibfnamefont {M.}~\bibnamefont
  {Geilen}}, \bibinfo {author} {\bibfnamefont {F.}~\bibnamefont {Kohl}},
  \bibinfo {author} {\bibfnamefont {A.}~\bibnamefont {Nicoloiu}}, \bibinfo
  {author} {\bibfnamefont {A.}~\bibnamefont {M{\"{u}}ller}}, \bibinfo {author}
  {\bibfnamefont {B.}~\bibnamefont {Hillebrands}},\ and\ \bibinfo {author}
  {\bibfnamefont {P.}~\bibnamefont {Pirro}},\ }\href
  {https://doi.org/10.1063/5.0029308} {\bibfield  {journal} {\bibinfo
  {journal} {Applied Physics Letters}\ }\textbf {\bibinfo {volume} {117}},\
  \bibinfo {pages} {213501} (\bibinfo {year} {2020})}\BibitemShut {NoStop}%
\bibitem [{\citenamefont {Ciubotaru}\ \emph {et~al.}(2016)\citenamefont
  {Ciubotaru}, \citenamefont {Devolder}, \citenamefont {Manfrini},
  \citenamefont {Adelmann},\ and\ \citenamefont {Radu}}]{Ciubotaru2016}%
  \BibitemOpen
  \bibfield  {author} {\bibinfo {author} {\bibfnamefont {F.}~\bibnamefont
  {Ciubotaru}}, \bibinfo {author} {\bibfnamefont {T.}~\bibnamefont {Devolder}},
  \bibinfo {author} {\bibfnamefont {M.}~\bibnamefont {Manfrini}}, \bibinfo
  {author} {\bibfnamefont {C.}~\bibnamefont {Adelmann}},\ and\ \bibinfo
  {author} {\bibfnamefont {I.~P.}\ \bibnamefont {Radu}},\ }\bibfield  {journal}
  {\bibinfo  {journal} {Applied Physics Letters}\ }\textbf {\bibinfo {volume}
  {109}},\ \href {https://doi.org/10.1063/1.4955030} {10.1063/1.4955030}
  (\bibinfo {year} {2016}),\ \Eprint {https://arxiv.org/abs/1602.08091}
  {arXiv:1602.08091} \BibitemShut {NoStop}%
\bibitem [{\citenamefont {Duquesne}\ \emph {et~al.}(2019)\citenamefont
  {Duquesne}, \citenamefont {Rovillain}, \citenamefont {Hepburn}, \citenamefont
  {Eddrief}, \citenamefont {Atkinson}, \citenamefont {Anane}, \citenamefont
  {Ranchal},\ and\ \citenamefont {Marangolo}}]{Duquesne2019}%
  \BibitemOpen
  \bibfield  {author} {\bibinfo {author} {\bibfnamefont {J.~Y.}\ \bibnamefont
  {Duquesne}}, \bibinfo {author} {\bibfnamefont {P.}~\bibnamefont {Rovillain}},
  \bibinfo {author} {\bibfnamefont {C.}~\bibnamefont {Hepburn}}, \bibinfo
  {author} {\bibfnamefont {M.}~\bibnamefont {Eddrief}}, \bibinfo {author}
  {\bibfnamefont {P.}~\bibnamefont {Atkinson}}, \bibinfo {author}
  {\bibfnamefont {A.}~\bibnamefont {Anane}}, \bibinfo {author} {\bibfnamefont
  {R.}~\bibnamefont {Ranchal}},\ and\ \bibinfo {author} {\bibfnamefont
  {M.}~\bibnamefont {Marangolo}},\ }\href
  {https://doi.org/10.1103/PhysRevApplied.12.024042} {\bibfield  {journal}
  {\bibinfo  {journal} {Physical Review Applied}\ }\textbf {\bibinfo {volume}
  {12}},\ \bibinfo {pages} {1} (\bibinfo {year} {2019})}\BibitemShut {NoStop}%
\bibitem [{\citenamefont {Herring}\ and\ \citenamefont
  {Kittel}(1951)}]{Herring1951}%
  \BibitemOpen
  \bibfield  {author} {\bibinfo {author} {\bibfnamefont {C.}~\bibnamefont
  {Herring}}\ and\ \bibinfo {author} {\bibfnamefont {C.}~\bibnamefont
  {Kittel}},\ }\href {https://doi.org/10.1103/PhysRev.81.869} {\bibfield
  {journal} {\bibinfo  {journal} {Physical Review}\ }\textbf {\bibinfo {volume}
  {81}},\ \bibinfo {pages} {869} (\bibinfo {year} {1951})}\BibitemShut
  {NoStop}%
\bibitem [{\citenamefont {Vansteenkiste}\ \emph {et~al.}(2014)\citenamefont
  {Vansteenkiste}, \citenamefont {Leliaert}, \citenamefont {Dvornik},
  \citenamefont {Helsen}, \citenamefont {Garcia-Sanchez},\ and\ \citenamefont
  {{Van Waeyenberge}}}]{Vansteenkiste2014}%
  \BibitemOpen
  \bibfield  {author} {\bibinfo {author} {\bibfnamefont {A.}~\bibnamefont
  {Vansteenkiste}}, \bibinfo {author} {\bibfnamefont {J.}~\bibnamefont
  {Leliaert}}, \bibinfo {author} {\bibfnamefont {M.}~\bibnamefont {Dvornik}},
  \bibinfo {author} {\bibfnamefont {M.}~\bibnamefont {Helsen}}, \bibinfo
  {author} {\bibfnamefont {F.}~\bibnamefont {Garcia-Sanchez}},\ and\ \bibinfo
  {author} {\bibfnamefont {B.}~\bibnamefont {{Van Waeyenberge}}},\ }\href
  {https://doi.org/10.1063/1.4899186} {\bibfield  {journal} {\bibinfo
  {journal} {AIP Advances}\ }\textbf {\bibinfo {volume} {4}},\ \bibinfo {pages}
  {107133} (\bibinfo {year} {2014})},\ \Eprint
  {https://arxiv.org/abs/1406.7635} {arXiv:1406.7635} \BibitemShut {NoStop}%
\bibitem [{ait()}]{aithericon}%
  \BibitemOpen
  \href {https://aithericon.com} {\bibinfo {title}
  {aithericon.com}}\BibitemShut {NoStop}%
\bibitem [{\citenamefont {Vanderveken}\ \emph {et~al.}(2021)\citenamefont
  {Vanderveken}, \citenamefont {Mulkers}, \citenamefont {Leliaert},
  \citenamefont {{Van Waeyenberge}}, \citenamefont {Sor{\'{e}}e}, \citenamefont
  {Zografos}, \citenamefont {Ciubotaru},\ and\ \citenamefont
  {Adelmann}}]{Vanderveken2021}%
  \BibitemOpen
  \bibfield  {author} {\bibinfo {author} {\bibfnamefont {F.}~\bibnamefont
  {Vanderveken}}, \bibinfo {author} {\bibfnamefont {J.}~\bibnamefont
  {Mulkers}}, \bibinfo {author} {\bibfnamefont {J.}~\bibnamefont {Leliaert}},
  \bibinfo {author} {\bibfnamefont {B.}~\bibnamefont {{Van Waeyenberge}}},
  \bibinfo {author} {\bibfnamefont {B.}~\bibnamefont {Sor{\'{e}}e}}, \bibinfo
  {author} {\bibfnamefont {O.}~\bibnamefont {Zografos}}, \bibinfo {author}
  {\bibfnamefont {F.}~\bibnamefont {Ciubotaru}},\ and\ \bibinfo {author}
  {\bibfnamefont {C.}~\bibnamefont {Adelmann}},\ }\href
  {https://doi.org/10.1103/PhysRevB.103.054439} {\bibfield  {journal} {\bibinfo
   {journal} {Physical Review B}\ }\textbf {\bibinfo {volume} {103}},\ \bibinfo
  {pages} {054439} (\bibinfo {year} {2021})},\ \Eprint
  {https://arxiv.org/abs/2011.10326} {arXiv:2011.10326} \BibitemShut {NoStop}%
\bibitem [{\citenamefont {Gueye}\ \emph {et~al.}(2016)\citenamefont {Gueye},
  \citenamefont {Zighem}, \citenamefont {Belmeguenai}, \citenamefont {Gabor},
  \citenamefont {Tiusan},\ and\ \citenamefont {Faurie}}]{Gueye2016}%
  \BibitemOpen
  \bibfield  {author} {\bibinfo {author} {\bibfnamefont {M.}~\bibnamefont
  {Gueye}}, \bibinfo {author} {\bibfnamefont {F.}~\bibnamefont {Zighem}},
  \bibinfo {author} {\bibfnamefont {M.}~\bibnamefont {Belmeguenai}}, \bibinfo
  {author} {\bibfnamefont {M.~S.}\ \bibnamefont {Gabor}}, \bibinfo {author}
  {\bibfnamefont {C.}~\bibnamefont {Tiusan}},\ and\ \bibinfo {author}
  {\bibfnamefont {D.}~\bibnamefont {Faurie}},\ }\bibfield  {journal} {\bibinfo
  {journal} {Journal of Physics D: Applied Physics}\ }\textbf {\bibinfo
  {volume} {49}},\ \href {https://doi.org/10.1088/0022-3727/49/14/145003}
  {10.1088/0022-3727/49/14/145003} (\bibinfo {year} {2016})\BibitemShut
  {NoStop}%
\bibitem [{\citenamefont {Kittel}(1949)}]{RevModPhys.21.541}%
  \BibitemOpen
  \bibfield  {author} {\bibinfo {author} {\bibfnamefont {C.}~\bibnamefont
  {Kittel}},\ }\href {https://doi.org/10.1103/RevModPhys.21.541} {\bibfield
  {journal} {\bibinfo  {journal} {Rev. Mod. Phys.}\ }\textbf {\bibinfo {volume}
  {21}},\ \bibinfo {pages} {541} (\bibinfo {year} {1949})}\BibitemShut
  {NoStop}%
\bibitem [{\citenamefont {Duflou}\ \emph {et~al.}(2017)\citenamefont {Duflou},
  \citenamefont {Ciubotaru}, \citenamefont {Vaysset}, \citenamefont {Heyns},
  \citenamefont {Sor{\'{e}}e}, \citenamefont {Radu},\ and\ \citenamefont
  {Adelmann}}]{Duflou2017}%
  \BibitemOpen
  \bibfield  {author} {\bibinfo {author} {\bibfnamefont {R.}~\bibnamefont
  {Duflou}}, \bibinfo {author} {\bibfnamefont {F.}~\bibnamefont {Ciubotaru}},
  \bibinfo {author} {\bibfnamefont {A.}~\bibnamefont {Vaysset}}, \bibinfo
  {author} {\bibfnamefont {M.}~\bibnamefont {Heyns}}, \bibinfo {author}
  {\bibfnamefont {B.}~\bibnamefont {Sor{\'{e}}e}}, \bibinfo {author}
  {\bibfnamefont {I.~P.}\ \bibnamefont {Radu}},\ and\ \bibinfo {author}
  {\bibfnamefont {C.}~\bibnamefont {Adelmann}},\ }\bibfield  {journal}
  {\bibinfo  {journal} {Applied Physics Letters}\ }\textbf {\bibinfo {volume}
  {111}},\ \href {https://doi.org/10.1063/1.5001077} {10.1063/1.5001077}
  (\bibinfo {year} {2017}),\ \Eprint {https://arxiv.org/abs/1708.06428}
  {arXiv:1708.06428} \BibitemShut {NoStop}%
\bibitem [{\citenamefont {M{\"{u}}ller}\ \emph {et~al.}(2015)\citenamefont
  {M{\"{u}}ller}, \citenamefont {Giangu}, \citenamefont {Stavrinidis},
  \citenamefont {Stefanescu}, \citenamefont {Stavrinidis}, \citenamefont
  {Dinescu},\ and\ \citenamefont {Konstantinidis}}]{Muller2015}%
  \BibitemOpen
  \bibfield  {author} {\bibinfo {author} {\bibfnamefont {A.}~\bibnamefont
  {M{\"{u}}ller}}, \bibinfo {author} {\bibfnamefont {I.}~\bibnamefont
  {Giangu}}, \bibinfo {author} {\bibfnamefont {A.}~\bibnamefont {Stavrinidis}},
  \bibinfo {author} {\bibfnamefont {A.}~\bibnamefont {Stefanescu}}, \bibinfo
  {author} {\bibfnamefont {G.}~\bibnamefont {Stavrinidis}}, \bibinfo {author}
  {\bibfnamefont {A.}~\bibnamefont {Dinescu}},\ and\ \bibinfo {author}
  {\bibfnamefont {G.}~\bibnamefont {Konstantinidis}},\ }\href
  {https://doi.org/10.1109/LED.2015.2494363} {\bibfield  {journal} {\bibinfo
  {journal} {IEEE Electron Device Letters}\ }\textbf {\bibinfo {volume} {36}},\
  \bibinfo {pages} {1299} (\bibinfo {year} {2015})}\BibitemShut {NoStop}%
\end{thebibliography}%
	\clearpage
	\appendix
	
	\section{Analysis of anisotropy}
\begin{figure}[hb]
		\includegraphics[width=8cm]{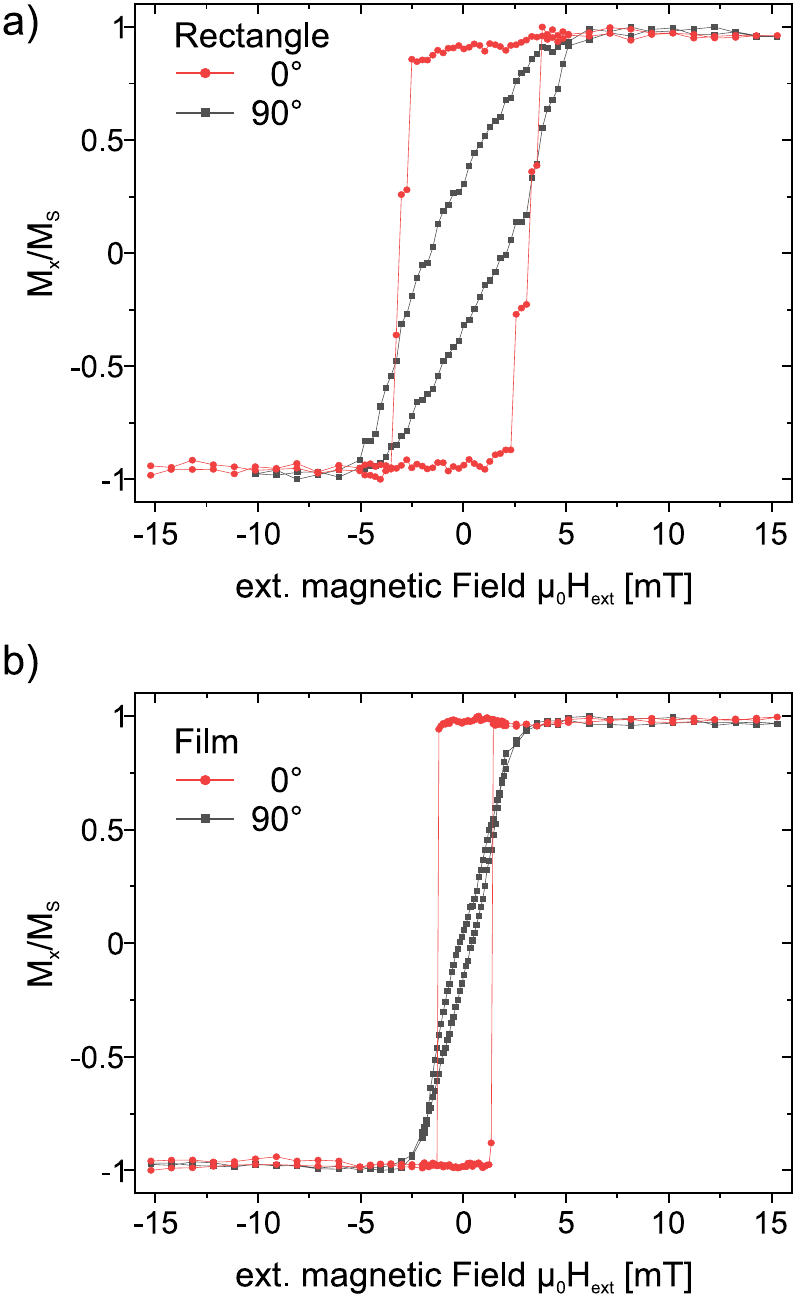}
		\centering
		\caption{Hysteresis loops for a) a rectangle and b) a film with dimensions as described in Fig. \ref{fig:fig1} measured by Kerr microscopy. $\varphi =$ 0° corresponds to the long axis of the rectangle.  }
		\label{fig:Append1}
	\end{figure}

\begin{figure}[ht]
		\includegraphics[width=8cm]{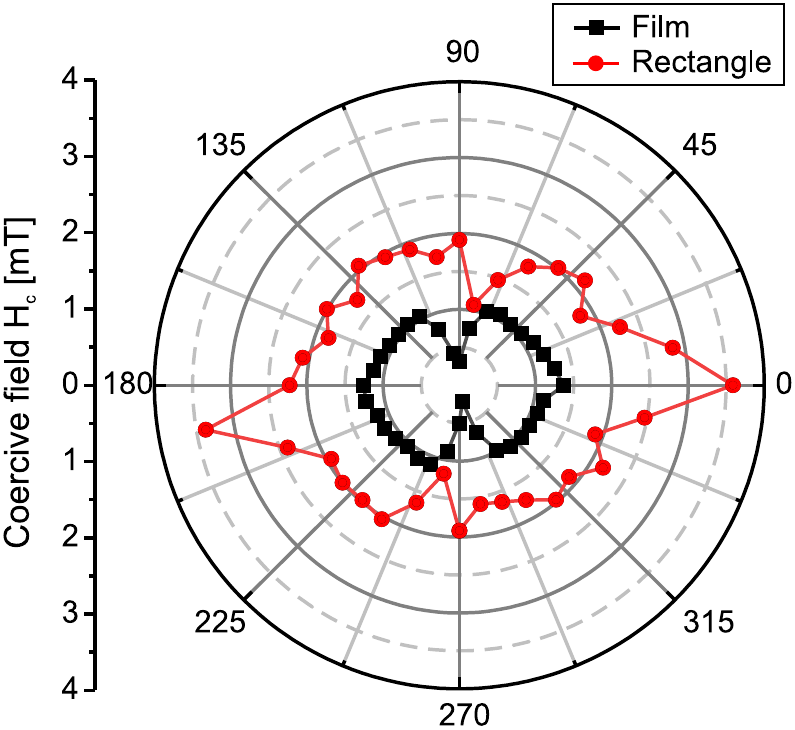}
		\centering
		\caption{Coercive field $H_\mathrm{c}$ as function of the angle $\varphi$. Film and rectangle show a unidirectional anisotropy along the x-axis.}
		\label{fig:Append2}
	\end{figure}

	In the following, the determination of the anisotropy constant used in the micromagnetic simulations is discussed.\\
	By means of Kerr microscopy (longitudinal mode), the hysteresis curve was obtained for the rectangular CoFeB structure described in the manuscript, as well as for a square with $\unit[350]{\upmu m}$ edge length.  This larger structure can be assumed to be an infinitely extended film. To determine the anisotropy, the hysteresis was measured for different angles $\varphi$. Figure \ref{fig:Append1} shows the two hysteresis curves for the extreme cases of the hard and the easy axis. 
	The coercive field $H_\mathrm{c}$ was taken from the hysteresis curves. As can be seen in Fig.~\ref{fig:Append2}, both the rectangle and the film show an easy anisotropy axis lying along the long axis of the rectangle. The anisotropy of the rectangle is more pronounced than for the film, since the shape anisotropy and the growth-induced anisotropy add up in this case. Using the Stoner-Wohlfarth model, a lower bound for the anisotropic constant can be estimated:
	\begin{equation}
		K_u = 2\upmu_0 H_\mathrm{c} M_\mathrm{S}.
	\end{equation}\\
	This yields an anisotropy constant for the growth-induced anisotropy of $K_\mathrm{u}=\unit[755]{J/m^3}$.
	Since this is only a lower limit due to the formation of domain walls, which the model does not take into account, the value of $K_\mathrm{u} =\unit[1600]{J/m^3}$ was adjusted to the measured values by means of micromagnetic simulations as follows. For the simulation, an excitation frequency of $f=\unit[6.3]{GHz}$ was chosen and the field $\upmu_0 H_\mathrm{ext}=\unit[6.4]{mT}$, which agrees with the experimental value. 
	\begin{figure}[hb!]
		\includegraphics[width=8cm]{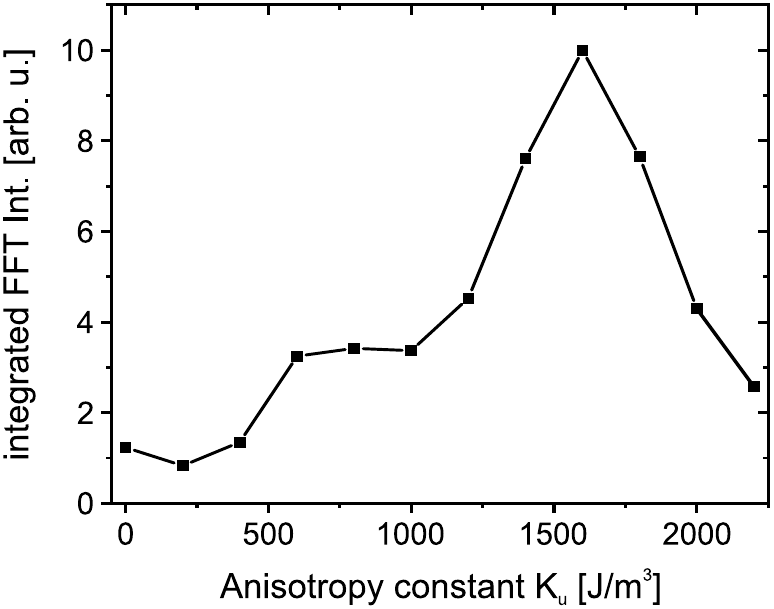}
		\centering
		\caption{Integrated FFT intensity as function of the anisotropy constant $K_\mathrm{u}$. The maximum was chosen for the simulations shown in the manuscript. }
		\label{fig:Append3}
	\end{figure}
	An elastic wave, as described in the manuscript, was used for excitation, while the anisotropy constant is varied for uniaxial anisotropy. The evaluation is also performed as described above. Figure \ref{fig:Append3} shows the integrated FFT intensity obtained as a function of the anisotropy constant.
	
	\section{Magneto-elastic interaction in micromagnetic simulations}
\begin{figure}[hbt]
		\includegraphics[width=8cm]{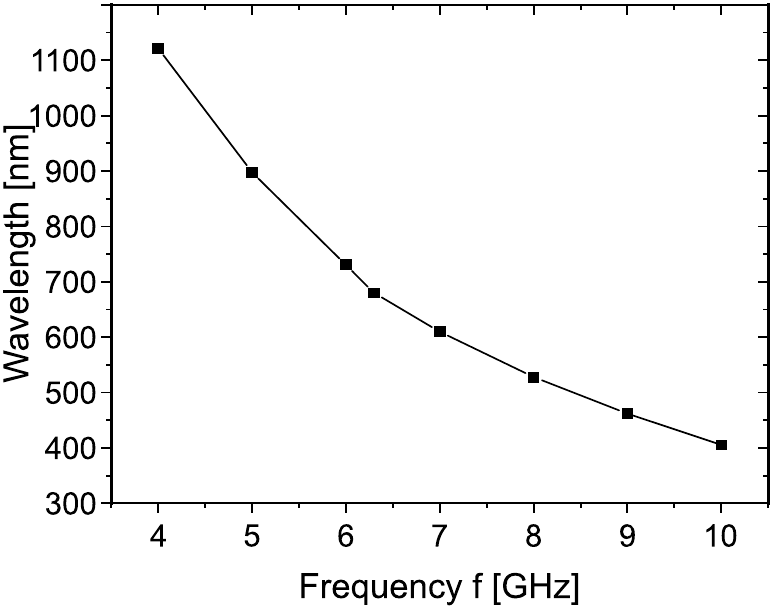}
		\centering
		\caption{Wavelength of the SAW as function of the frequency.}
		\label{fig:Append4}
	\end{figure}

	In the main text, the modeling of elastic waves in MuMax3 is discussed. A plane wave is assumed for the individual strain components. MuMax3 calculates the resulting magneto-elastic field for each time step and cell according to equation \ref{eq:MEL} \cite{RevModPhys.21.541,Duflou2017}:
	\begin{equation}
		\label{eq:MEL}
		\upmu_0\mathbf{H}^\mathrm{mel}= - \frac{1}{M_S^2}\left(\begin{array}{cc}
			2B_1 M_\mathrm{x} S_\mathrm{xx} + B_2 (M_\mathrm{y} S_\mathrm{xy} + M_\mathrm{z} S_\mathrm{xz})  \\
			2B_1 M_\mathrm{y} S_\mathrm{yy} + B_2 (M_\mathrm{x} S_\mathrm{yx} + M_\mathrm{z} S_\mathrm{yz})  \\
			2B_1 M_\mathrm{z} S_\mathrm{zz} + B_2 (M_\mathrm{x} S_\mathrm{zx} + M_\mathrm{y} S_\mathrm{zy}) 
		\end{array}\right).
	\end{equation}
	Furthermore, it was pointed out that the wavelength of the SAWs changes with frequency $f$. The values shown in Figure~\ref{fig:Append4} are taken from Ref. \cite{Muller2015} and Ref. \cite{Geilen2020}.

\end{document}